\documentstyle[12pt]{article}

\def\bB{${\partial {\cal B}}$}
\def\iB{${\cal B}$}
\def\dE{$<|\Delta E|>$}
\begin{document}

\begin{flushright}
Preprint CAMTP/95-5\\
September 1995\\
\end{flushright}

\begin{center}
\large
{\bf Is there relevance of chaos in numerical solutions of quantum billiards?}
\\
\vspace{0.3in}
\normalsize
Baowen Li\footnote{e-mail Baowen.Li@UNI-MB.SI}
 and Marko Robnik\footnote{e-mail Robnik@UNI-MB.SI}\\

\vspace{0.2in}
Center for Applied Mathematics and Theoretical Physics,\\
University of Maribor, Krekova 2, SLO-62000 Maribor, Slovenia\\
\end{center}
\vspace{0.2in}
\normalsize
{\bf Abstract.}
In numerically solving the Helmholtz equation inside a connected plane domain
with Dirichlet boundary conditions (the problem of the quantum billiard) one
surprisingly faces enormous difficulties if the domain has a problematic
geometry such as various nonconvex shapes. We have tested several general
numerical methods in solving the quantum billiards. Following our previous
paper (Li and Robnik 1995) where we analyzed the Boundary Integral Method
(BIM), in the present paper we investigate systematically the so-called Plane
Wave Decomposition Method (PWDM) introduced and advocated by Heller (1984,
1991).  In contradistinction to BIM we find that in PWDM the classical chaos is
definitely relevant for the numerical accuracy at fixed density of
discretization on the boundary $b$ ($b$ = number of numerical nodes on the
boundary within one de Broglie wavelength). This can be understood
qualitatively and is illustrated for three one-parameter families of billiards,
namely Robnik billiard, Bunimovich stadium and Sinai billiard.  We present
evidence that it is not only the ergodicity which matters, but also the
Lyapunov exponents and Kolmogorov entropy. Although we have no quantitative
theory we believe that this phenomenon is one  manifestation of quantum chaos.
\\\\
Submitted to {\bf Journal of Physics A}

\normalsize
\vspace{0.3in}
\newpage

\section{Introduction}
It is quite embarrassing to realize that in an attempt to numerically solve the
Helmholtz equation
\begin{equation}
\nabla_{{\bf r}}^2 \psi({\bf r}) + k^2 \psi({\bf r}) = 0,
\label{eq:Helmholtz}
\end{equation}
satisfied by the scalar solution $\psi({\bf r})$ with eigenenergy $E=k^2$
inside  a connected plane domain ${\cal B}$ with the Dirichlet boundary
condition $\psi({\bf r})=0$ on the boundary ${\partial {\cal B}}$, one can face
enormous difficulties in cases of "problematic" geometries such as e.g. various
nonconvex shapes. This is precisely the problem of solving and describing the
quantum billiard ${\cal B}$ as a Hamiltonian dynamical system, which is thus
just the 2-dim Schr\"odinger problem for a free point particle moving inside
the
enclosure ${\partial {\cal B}}$, described by the wavefunction $\psi({\bf r})$
with the eigenenergy $E=k^2$. The corresponding classical problem is the
classical dynamics of a freely moving point particle obeying the law of
specular
reflection upon hitting the boundary ${\partial{\cal B}}$.
Quantum billiards and their correspondence to their classical counterparts,
especially in the semiclassical level, are important model systems in studies
of
quantum chaos (Gutzwiller 1990, Giannoni {\em et al} 1991).
There are several {\em general} methods for a numerical solution of
equation (\ref{eq:Helmholtz}) such as the Boundary Integral Method (BIM),
recently carefully and extensively studied by Li and Robnik (1995a), following
the important paper by Berry and Wilkinson (1984) (see also Boasman 1994),
and the Plane Wave Decomposition Method (PWDM), introduced and advocated by
Heller
(1984,1991), whose analysis --- especially in the light of the relevance of
classical chaos --- is the subject of our present paper. Another quite general
method is the conformal mapping diagonalization technique introduced by Robnik
(1984) and further developed by Berry and Robnik (1986) and recently by Prosen
and Robnik (1993,1994), which in principle works for any shape whereas in
practice it is used for shapes for which the conformal mapping onto the unit
disk\footnote{or some other integrable geometries admitting a simple basis for
the representation}
is sufficiently simple (possiblly also analytic). These methods can
face quite similar problems in cases of almost intractable geometries, but they
are to some extent complementary. For example, the conformal mapping
diagonalization technique can
provide a complete set of all eigenenergies up to some maximal value beyond
which the calculations cannot be performed due to the lack of computer storage
(RAM),  which means that we cannot reach very high-lying eigenstates.
(Our present record (Prosen and Robnik 1994) is about 35,000 for the size of
the banded matrix that we diagonalize in double precision, yielding at least
12,000 good levels with accuracy of at least one percent of the mean level
spacing.)

However, using PWDM it is possible to go higher in energy by orders of
magnitude but then only a few selected states can be calculated with many
intermediate states in the spectral stretch missing. Therefore the geometry of
some interesting and representative high-lying states can be analyzed,  but the
sample is typically not sufficiently complete (many missing states) to perform
statistical analysis. See e.g. our recent papers on this topics (Li and Robnik
1994a, 1995b, 1995c, 1995d). The reasons for a failure of one of these methods
can be quite different. For example, in BIM the main difficulty stems from the
existence of "exterior chords" in nonconvex geometries in its standard
formulation, but the trouble might be overcome by an appropriate reformulation
of the method adapted to the correct semiclassical behaviour. This is discussed
in  (Li and Robnik 1995a) where we also show that classical chaos is completely
irrelevant for BIM. On the contrary, in PWDM we find that the classical chaos
is relevant for numerical accuracy especially in the semiclassical limit of
sufficiently small effective Planck constant $\hbar_{eff}$
reached at sufficiently
high eigenenergies. This demonstration and its qualitative explanation is the
main subject of our present paper. To give a specific example we should mention
isospectral billiards discovered and proved by Gordon {\em et al} (1992)  which
have been investigated experimentally (Sridhar and Kudrolli 1994) and it is
also our experience (Li and Robnik 1994b) that BIM fails completely in this
case (namely, due to strong nonconvexities) whereas  PWDM at $b=12$ yields the
accuracy of eigenenergies of about within a few percent of the mean level
spacing, except for some very special eigenmodes for which surprisingly
we find agreement
within double precision (16 digits) and which are characterized by the fact
that these eigenvalues agree with the analytic solutions for the triangles
within single precision (8 digits). So the fact that in this and similar cases
the experimental precision (for some levels) exceeds the best possible
numerical precision even when using the best available methods is embarrassing
for a theoretician but also motivation for further work.

\section{The plane wave decomposition method of Heller}
In this section we present our general exposition of PWDM following Li and
Robnik (1994a). To solve the Schr\"odinger equation  (\ref{eq:Helmholtz})
for $\psi({\bf r})$ with Dirichlet boundary condition $\psi({\bf r}) =0$ on
\bB $\;$ we use the {\em Ansatz} of the following superposition  of plane waves
(originally due to Heller (1984))
\begin{equation}
\psi({\bf r}) = \sum_{j=1}^{N} a_{j} \cos (k_{xj}x + k_{yj}y + \phi_j),
\label{eq:pwd}
\end{equation}
where $k_{xj} = k\cos\theta_j,\quad k_{yj} = k\sin\theta_j,\quad k^2 = E$,
and we use the notation ${\bf r} = (x,y)$. $N$ is the number of plane waves
and $\phi_j$ are {\em random phases}, drawn from the
interval $[0,2\pi)$, assuming
uniform distribution, and $\theta_j=2j\pi/N$ determining the direction angles
of
the wavevectors chosen equidistantly. The {\em Ansatz} (\ref{eq:pwd}) solves
the Schr\"odinger equation (\ref{eq:Helmholtz}) in the interior of the billiard
region \iB, so that we have only to satisfy the Dirichlet boundary condition.
Taking the random phases, as we discovered, is equivalent
to spreading the origins of plane waves all over the billiard region,
and at the same time this results in reducing the CPU-time by
almost a factor of ten.
For a given $k$ we put the wavefunction equal to zero at
a finite number $M$ of boundary points (primary nodes) and
equal to 1 at an arbitrarily chosen interior point. Of course, $M\ge N$.
This gives an inhomogeneous set of equations which can be solved
by matrix inversion.  Usually the matrix is very singular,  thus the
{\em Singular Value Decomposition} (SVD) method has been invoked
(Heller 1984, Press {\em et al} 1986).
After obtaining the coefficients $a_{j}$ we
calculate the wavefunctions at other boundary points (secondary nodes).
We always have 3 secondary nodes between a pair of primary nodes. The
experience shows that further increase of the number of secondary nodes does
not enhance the accuracy.
The sum of the squares of the wavefunction at all the secondary nodes
(Heller called this sum "tension") would  be ideally zero if $k^2$ is an
eigenvalue and if (\ref{eq:pwd}) is the corresponding exact solution of
(\ref{eq:Helmholtz}).
In practice it is a positive number. Therefore the eigenvalue
problem now is to find the minimum of the "tension". In our numerical
procedure we have looked for the zeros of
the first derivative of the tension; namely the derivative is available
analytically/explicitly from (\ref{eq:pwd}) once the amplitudes $a_j$
have been found.  In this paper we make the choice $M=N$, since it proved
to be sufficient for calculating the lowest 100 states whose accuracy we
analyze. (For high-lying states studied in (Li and Robnik 1994a) we have used
$M = 5N/3$.)
It must be pointed out that the wavefunctions obtained in
this way are not (yet) normalized, due to the arbitrary choice of the
interior point where the value of the wavefunction has been arbitrarily
set equal to unity. We therefore explicitly normalize these wave functions
before embarking to the analysis of their properties.
\\\\
The accuracy of this method of course depends on the number of plane waves
($N$) and on the  number of the primary nodes ($M$), and we have a considerable
freedom in choosing $N$ and $M\ge N$.
In order to reach a sufficient accuracy the experience shows that
we should take at least $N=3{\cal L}/\lambda_{de Broglie}$, and $M=N$, where
${\cal L}$ is the perimeter of the billiard and $\lambda_{de Broglie}$
is the de Broglie wavelength $=2\pi/k$. With this choice in the present context
and for the lowest 100 states we reach the double
precision accuracy (sixteen digits) for all levels of integrable
systems like rectangular billiard (where the eigenenergies can be
given trivially analytically) and the circular billiard, but also
for Robnik billiard  ${\cal B}_{\lambda}$ for small $\lambda \le 0.1$.
Introducing the density of discretization $b$ defined as the number of
numerical nodes per one de Broglie wavelength on the boundary we thus write the
number of plane waves $N = b {\cal L}/\lambda_{de Broglie} = b2\pi{\cal L}/k$.
\\\\
The main problem of investigation in this paper is to study the dependence of
the systematical numerical error $\Delta E$ (i.e. the error due to the finite
discretization) on the density of discretization
$b$, and the dependence of $\Delta E$ on the geometry (billiard shape
parameter) at fixed $b$. In order to perform a systematic analysis the errors
should be measured in some natural units and in our case this is of course just
the mean level spacing which according to the leading term of Weyl formula is
equal to $4\pi/{\cal A}$ where ${\cal A}$ is the area of the billiard \iB. From
now on we shall always assume that $\Delta E$ of a particular energy level is
in fact measured in such natural units.
Of course one immediately realizes that the error $\Delta E$ fluctuates
wildly from state to state (see figure 5)
so that generally nothing can be predicted about it individually.
Therefore the approach must be a statistical one and so we typically
take an average of the errors $\Delta E$ over a suitable ensemble of states.
Specifically, in all cases of this paper we have taken the average of the
absolute values of $\Delta E$ over the lowest 100 states (of a given symmetry
class) and denote it by $<|\Delta E|>$.
It is important and should be mentioned that we have also
checked the stationarity of such average value over consecutive
spectral stretches of 100 states each, so that our procedure does make sense.
See a discussion at the end of section 3.
\\\\
It turns out that the accuracy of energy levels depends nontrivially on $b$,
unlike in BIM (where we find always a power law (Li and Robnik 1995a)), namely
it typically shows broken power law. By this we mean that\\
\dE obeys a power
law \dE$=Ab^{-\alpha}$ with very large $\alpha$ for sufficiently small $b$,
$b\le b_c$, whereas for larger $b\ge b_c$ it obeys a rather flat power law with
very small positive $\alpha$ (close to zero). Therefore in contradistinction
to BIM it
is difficult to explore the general dependence of \dE on $b$, if there is any
such universality at all. However, in order to investigate the dependence of
the accuracy on geometry and the implied dynamical properties of billiards,
we have
decided to fix the value of $b$ and  have chosen $b=12$, and then we look at
the dependence of \dE on the shape parameters of three one-parameter billiards,
namely Robnik billiard, Bunimovich stadium and Sinai billiard.

\section{Numerical results}

As is well known in classically integrable quantum Hamiltonian systems in the
semiclassical limit (of sufficiently small $\hbar$)
the eigenfunction can locally be described by a {\em finite}
superposition of plane waves with the same wavenumber, in case of plane
billiards it is  $k=\sqrt{E}$. If the quantum system has ergodic classical
dynamics then in the semiclassical limit locally the wavefunction can be
represented as a superposition of {\em infinitely} many plane waves with the
same $k$ and with the wavevectors being isotropically distributed on the
circle of radius $k$. Moreover, the ergodicity suggests to assume random phases
for the ensemble of plane waves which implies that to the lowest approximation
the wavefunction is a Gaussian random function. While this is a good starting
approximation, originally due to Berry (1977) and recently verified by Aurich
and Steiner (1993) and also by Li and Robnik (1994a), the phases are actually
not random but correlated in a subtle way dictated by the classical dynamics
especially along the short and the least unstable periodic orbits, which is the
origin of the scar phenomenon (Heller 1984, 1991, Bogomolny 1988, Berry 1989,
Robnik 1989, Li and Robnik 1995d). (For a discussion see (Robnik 1988, 1995).)
Thus we can qualitatively very well
understand that PWDM should work well or even brilliantly in cases of
classically integrable billiards whereas in the ergodic systems we expect a
severe degradation of the accuracy (at fixed $b$) simply because the finite
number of plane waves cannot capture the correct (infinite) superposition of
plane waves everywhere in the interior of the billiard. If the system is
a generic system of a mixed type with regular and irregular regions coexisting
in the classical phase space, a scenario described by the KAM theory, then the
degradation of accuracy (at fixed $b$) with increasing fractional measure of
the chaotic component (denoted by $\rho_2$) is certainly expected. However,
$\rho_2$ is not the only parameter which controls the accuracy (at fixed $b$),
since, as we shall see, the dynamical properties like diffusion time, Lyapunov
exponent and the Kolmogorov entropy play a role. It is the aim of the
present paper to numerically explore this type of behaviour in three different
billiard systems.
\\\\
The first billiard system is defined as the quadratic (complex) conformal map
$w = z + \lambda z^2$ from the unit disk $|z|\le 1$ from the $z$ plane onto the
$w = (x,y)$ complex plane. The system has been introduced by Robnik (1983) and
further studied by Hayli {\em et al} (1987), Frisk (1990), Bruus and Stone
(1994) and Stone and Bruus (1993a,b) for various parameter values $\lambda$.
Since the billiard (usually called Robnik billiard) has analytic boundary it
goes continuously from integrable case (circle, $\lambda=0$) through a KAM-like
regime of small $\lambda\le 1/4$ with mixed classical dynamics, becomes
nonconvex at $\lambda=1/4$ (the bounce map becomes discontinuous), where the
Lazutkin caustics (invariant tori) are destroyed giving way to ergodicity. As
shown by Robnik (1983) the classical dynamics at these values of $\lambda$ is
predominantly chaotic (almost ergodic), although Hayli {\em et al} (1987) have
shown that there are still some stable periodic orbits surrounded by very tiny
stability islands up to $\lambda=0.2791$. At larger $\lambda$ we have
reason and numerical evidence (Li and Robnik 1994c) to expect that the
dynamics can be ergodic. It has been recently rigorously proven by Markarian
(1993) that for $\lambda =1/2$ (cardioid billiard) the system is indeed
ergodic, mixing and K. This was a further motivation to study the cardioid
billiard classically, semiclassically and quantanlly by several groups e.g. by
B\"acker {\em et al} (1994) and B\"acker (1995), Bruus and Whelan (1995).
The billiard shape for
$\lambda =0.4$ is  shown (the upper half) in figure 1a. Since all states are
either even or odd we can take into account these symmetry properties
explicitly. In fact we want to specialize to the odd eigenstates only.
Therefore in order to a priori satisfy the Dirichlet boundary condition on the
abscissa of figure 1a we specialize the general {\em Ansatz} (\ref{eq:pwd}) to
the following form
\begin{equation}
\psi({\bf r}) = \sum_{j=1}^{N} a_j \cos (k_{xj}x + \phi_j) \sin(k_{yj}y),
\label{eq:robnik}
\end{equation}
where all the quantities are precisely as in (\ref{eq:pwd}) except that the
$N$ discretization (primary) nodes are equidistantly located only
along the half of the full billiard boundary, so that $b$ is exactly the same
as in using the {\em Ansatz} (\ref{eq:pwd}) for the full billiard.
\\\\
In figure 2 we show the results for this billiard, namely we plot \dE versus
$\lambda$ at fixed $b=12$. Close to integrability ($\lambda\le0.1$) we reach
the accuracy within 14 to 15 digits which is almost the double precision on
our machine (16 digits), in which all our calculations have been performed.
As the value of $\lambda$ increases we observe a dramatic deterioration of the
accuracy where \dE increases by many orders of magnitude, namely almost by
13 decades, leveling off at \dE approximately equal to $10^{-2}$, which means
that we have now the accuracy of only a few percent of the mean level spacing.
This dramatic but quite smooth increase of \dE is certainly related to the
emergence of classical chaos with increasing $\lambda$, but definitely is
{\em not} controlled merely by $\rho_2$, because $\rho_2$ reaches the value of
1 (almost ergodicity) already at $\lambda =1/4$ (Robnik 1983, Prosen and Robnik
1993), whereas \dE still varies considerably in the region $\lambda \ge 1/4$.
Thus it is obvious that in the semiclassical picture also other classical
dynamical properties (measures of the "hardness" of chaos) play an important
role. Although we do not have a quantitative theory yet one should observe that
according to (Robnik 1983) the Lyapunov exponent and Kolmogorov entropy ($h$)
vary also quite smoothly with $\lambda$, suggesting a speculation that there
might be a relation between \dE and $h$.
\\\\
Another demonstration of the effectivity of PWDM and its accuracy is displayed
in table 1 where we show the numerical value of the scalar product of two
consecutive normalized eigenstates, namely the ground state and the first
excited state, denoted by $O_{12}$, which ideally should be zero.
As we see here too the accuracy decreases (by orders of magnitude)
sharply but smoothly with increasing shape parameter $\lambda$.
\\\\
It is then interesting to similarly analyze an ergodic system such as stadium
of Bunimovich shown in figure 1b, where the shape parameter is $a/R$ and we
have looked at the results for $0\le a/R \le 10$. In fact for our purposes we
have chosen and fixed $R=1$ in all cases. Since this billiard is known
to be rigorously ergodic (and mixing and K) for any $a > 0$ in this case
$\rho_2$ is exactly 1 and constant.
We have calculated the lowest 100 energy levels of the odd-odd symmetry class.
Therefore in this case the general {\em Ansatz} (\ref{eq:pwd}) can be
specialized as follows
\begin{equation}
\psi({\bf r}) = \sum_{j=1}^{N} a_j \sin(k_{xj}x)\sin(k_{yj}y).
\label{eq:stadium}
\end{equation}
Here again the discretization (primary) nodes are only on the outer boundary of
the stadium with discretization density $b=12$.
{}From our plot in figure 2 we see that in the
integrable case of the circle ($a =0$) we again reach the accuracy within at
least 14 digits, but this brilliant accuracy at fixed $b=12$ deteriorates
almost discontinuously upon increasing $a$ and then \dE still increases by
about two orders of magnitude when $a$ goes from 0.1 to 10. It appears to us
that classical chaos is definitely relevant for the accuracy of the method
which might and should be explained by an appropriate theory in the
semiclassical level. As an observation we should mention that the Kolmogorov
entropy increases sharply with $a/R$ when $a/R$ goes from 0 to about 1, where
it reaches the maximum, and then decreases slowly (Benettin and Strelcyn 1978),
whereas our \dE increases monotonically. Thus if there is a relationship
between
\dE and Kolmogorov entropy it certainly is not a simple one.
\\\\
We have tested also another system with hard chaos, namely the Sinai billiard
sketched in figure 1c (desymmetrized). The system is known to be ergodic,
mixing and K. In calculating the 100 lowest energy levels of the desymmetrized
Sinai billiard we used the same specialized {\em Ansatz} as in
eq. (\ref{eq:robnik}), thereby taking into account explicitly the Dirichlet
boundary condition on the abscissa $y=0$. In this case $b$ is the density of
discretization of the equidistant nodes along the rest of the perimeter.
Similarly as in case of stadium we easily reach the double
precision of 16 digits in the limiting integrable case of zero radius $R=0$,
but this accuracy is almost instantly lost by increasing $R$ as seen in figure
3. The value of \dE levels off at about $10^{-3}$ to $10^{-2}$ for all $R$
between $0.025$ and $0.45$.
\\\\
As a final technical point we comment on the stationarity of \dE as a function
of energy, which has been confirmed for the Robnik billiard at $\lambda=0.27$
where the average value over consecutive spectral stretches over 100 states has
been found to be quite stable.
Specifically, to illustrate this finding we plot in figure 5 the absolute
values
of the lowest 400 consecutive eigenstates where one can see that the average
value over 100 consecutive states is quite stable indeed.
This is shown in table 2 for four intevals of 100 states each.

\section{Discussion and conclusions}
We believe that our present paper presents quite firm numerical
(phenomenological) evidence for the relevance of classical chaos for the
effectiveness of PWDM as a quantal numerical method to solve a quantum
billiard, which is manifested especially in the semiclassical limit and might
and should be explained in terms of an appropriate semiclassical theory.
Qualitatively the reasons for this phenomenon are explained in the first
paragraph of section 3. The parameter $\rho_2$ (the fractional volume of
the chaotic component(s)) definitely plays an important
role but is not the only aspect of classical chaos controlling the behavior of
the error \dE at fixed discretization density $b$. Namely even in rigorously
ergodic systems where $\rho_2$ is 1 the error \dE might be controlled by the
slow diffusion in the classical phase space (diffusive ergodic regime, soft
chaos). If the classical diffusion time is much longer than the break time
$t_{break}$ ($t_{break} = \hbar/D$, $D$ = mean energy level spacing) then the
quantal states will be strongly localized in spite of the formal ergodicity
(for a demonstration see (Li and Robnik 1995b)), and
therefore they mimic certain amount of regularity enabling a better accuracy of
PWDM, i.e. \dE is smaller than for completely extended chaotic high-lying
eigenstates where according to our experience somehow \dE typically saturates
at
about a few percent of the mean level spacing, even if we brutally increase
$b$ beyond any reasonable limits.
Indeed, as can be seen by comparison of figures 2, 3 and 4, in case of the
stadium
this saturation value of \dE is about $10^{-4}$, which is almost two orders of
magnitude smaller than in the Sinai billiard (figure 4)
and cardioid billiard (figure 2).  We think that this is due to the strong
localization of eigenstates in the stadium, which is very well known to
display an unusual abundance of scars (Heller 1984, Li and Robnik 1995d).

Thus our present work is a motivation for a semiclassical theory to explain
this aspect of quantum chaos which exhibits some algorithmical properties of
PWDM in applying it to quantum billiards with variety of classical dynamics.

\section*{Acknowledgments}

The financial support by the Ministry of Science
and Technology of the Republic of Slovenia is gratefully acknowledged.
We thank Mag. Vladimir Alkalaj, director of the National Supercomputer Center,
for the kind support in using the supercomputer facilities.

\vfill
\newpage
\section*{References}
Abramowitz, M and Stegun I A (eds) 1972 {\em Handbook of mathematical
functions} (New York: Dover)\\\\
Aurich R and Steiner F 1993 {\em Physica} {\bf D64} 185\\\\
B\"acker A, Steiner F and Stifter P 1994 {\em Preprint} DESY-94-213\\\\
B\"acker A 1995 {\em Diplomarbeit}, February 1995
Universit\"at Hamburg, II. Institut f\"ur
Theoretische Physik\\\\
Benettin G and Strelcyn J M 1978 {\em Phys. Rev.} {\bf A17} 773\\\\
Berry M C 1977 {\em J. Phys. A: Math. Gen.} {\bf 10} 2083\\\\
Berry M V and Robnik M 1986 {\em J. Phys. A: Math. Gen.} {\bf 19} 649\\\\
Berry M V and Wilkinson M 1984 {\em Proc. Roy. Soc. Lond.} {\bf A392} 15\\\\
Berry M V 1989 {\em Proc. Roy. Soc. London} {\bf A423} 219\\\\
Boasman P A 1994 {\em Nonlinearity} {\bf 7} 485\\\\
Bogomolny E B 1988 {\em Physica} {\bf D31} 169\\\\
Bohigas O, Boos\'e D, Egydio de Carvalho R and Marvulle V 1993 {\em
Nuclear Physics}{\bf A560} 197\\\\
Bruus H and Stone A D 1994 {\em Phys. Rev.} {\bf B50} 18275\\\\
Bruus H and Whelan N D 1995 {\em Preprint} Niels Bohr Institute, Copenhagen,
chao-dyn/9509005\\\\
Frisk H 1990 Nordita {\em Preprint}\\\\
Giannoni M-J, Voros J and Zinn-Justin eds. 1991 {\em Chaos and Quantum Systems}
(Amsterdam: North-Holland)\\\\
Gordon C, Webb D and Wolpert S 1992 {\em Bull. Am. Math. Soc.} {\bf 27} 134\\\\
Gutzwiller M C 1990 {\em Chaos in Classical and Quantum Mechanics} (New York:
Springer)\\\\
Hayli A, Dumont T, Moulin-Ollagier J and Strelcyn J M 1987 {\em J. Phys. A:
Math. Gen} {\bf 20} 3237\\\\
Heller E J 1984 {\em Phys. Rev. Lett} {\bf 53} 1515\\\\
Heller E J 1991  in {\em Chaos and Quantum Systems (Proc. NATO ASI Les Houches
Summer School)} eds M-J Giannoni, A Voros and J Zinn-Justin,
(Amsterdam: Elsevier) p547\\\\
Li Baowen and Robnik M 1994a {\em J. Phys. A: Math. Gen.} {\bf 27} 5509\\\\
Li Baowen and Robnik M 1994b {\em unpublished}\\\\
Li Baowen and Robnik M 1994c {\em unpublished}\\\\
Li Baowen and Robnik M 1995a {\em Preprint} CAMTP/95-3, chao-dyn/9507002,
submitted to {\em J. Phys. A: Math. Gen.} in July\\\\
Li Baowen and Robnik M 1995b {\em J. Phys. A: Math. Gen.} {\bf 28} 2799\\\\
Li Baowen and Robnik M 1995c {\em J. Phys. A: Math. Gen.} {\bf 28} 4843\\\\
Li Baowen and Robnik M 1995d to be submitted to {\em J. Phys. A: Math.
Gen.}\\\\
Markarian R 1993 {\em Nonlinearity} {\bf 6} 819\\\\
Prosen T and Robnik M 1993 {\em J. Phys. A: Math. Gen.} {\bf 26} 2371\\\\
Prosen T and Robnik M 1994  {\em J. Phys. A: Math. Gen.} {\bf 27} 8059\\\\
Robnik M 1983 {\em J. Phys. A: Math. Gen.} {\bf 16} 3971\\\\
Robnik M 1984 {\em J. Phys. A: Math. Gen.} {\bf 17} 1049\\\\
Robnik M 1988 in {\em Atomic Spectra and Collisions in External Fields} eds. K
T Taylor, M H Nayfeh and C W Clark (New York: Plenum) 265-274\\\\
Robnik M 1989 {\em Preprint} Institute of Theoretical Physics, University of
California Santa Barbara\\\\
Robnik M 1995 {\em Introduction to Quantum Chaos} Preprint CAMTP/95-4, to
appear in the Proc. of Summer School and Conference "Complexity and Chaotic
Dynamics of Nonlinear Systems", Xanthi, Greece, July 1995\\\\
Sridhar S and Kudrolli A 1994 {\em Phys. Rev. Lett.} {\bf 72} 2175\\\\
Stone A D and Bruus H 1993a {\em Physica } {\bf B189 } 43\\\\
Stone A D and Bruus H 1993b {\em Surface Sci.} {\bf 305} 490\\\\

\newpage
\section*{Tables}

\bigskip
\bigskip

{\bf Table 1.} The test of the orthogonality of the eigenstates. The scalar
product of two consecutive normalized wavefunctions $O_{12}$, namely the
ground state and the first excited state, for Robnik billiard
at different shape parameters.\vspace{15mm}\\

\begin{tabular}{|c|c|}\hline
&\\
$\lambda$ & $O_{12}$\\[2.5ex] \hline
0   & 2.0E$-$16\\[1ex]
0.1 & -5.0E$-$15\\[1ex]
0.2 & 7.8E$-$10\\[1ex]
0.3 & 4.8E$-$6 \\[1ex]
0.4 & 5.5E$-$4 \\[1ex]
0.5 & 1.5E$-$3\\[1ex]
 \hline
\end{tabular}

\newpage

\bigskip
\bigskip
{\bf Table 2.}  The stationarity test of \dE for Robnik billiard at
$\lambda=0.27$ for the lowest 400 odd eigenstates. \vspace{15mm}\\

\begin{tabular}{|c|c|}\hline
&\\
average stretch & \dE\\[2.5ex] \hline
1-100   & 1.54E$-$7\\[1ex]
101-200 & 2.21E$-$7\\[1ex]
201-300 & 2.77E$-$7\\[1ex]
301-400 & 2.03E$-$7 \\[1ex]\hline
1-400 & 2.13E$-$7 \\[1ex]
\hline
\end{tabular}

\newpage
\section*{Figure captions}

\bigskip
\bigskip

\noindent {\bf Figure 1 (a-c):} The geometry of the boundary of the three
desymmetrized billiards, the Robnik billiard (a), the
Bunimovich stadium (b) and the Sinai billiard (c).

\bigskip
\bigskip

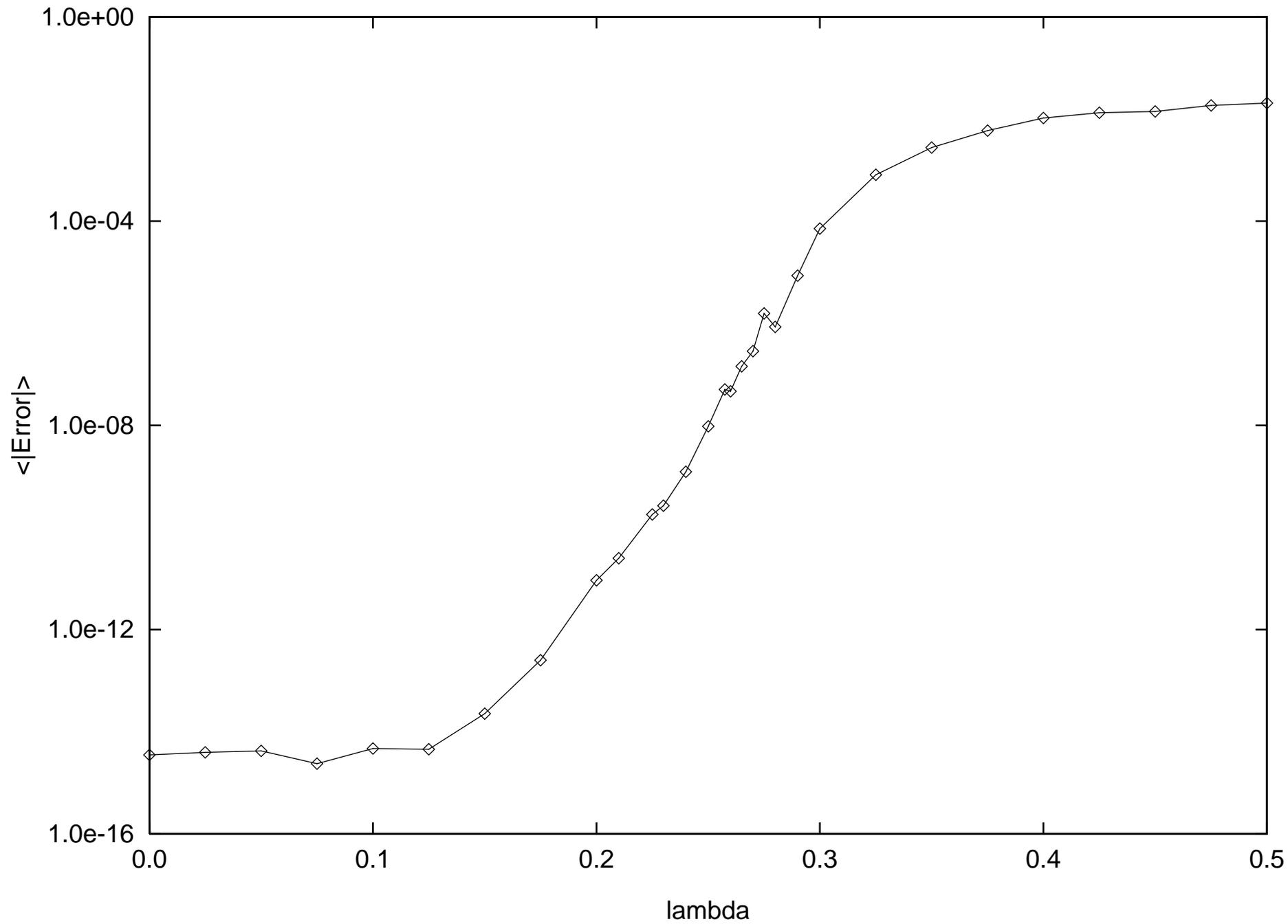
\noindent {\bf Figure 2:}
The ensemble averaged (over 100 lowest odd eigenstates) absolute
error (measured in units of mean level spacing) versus the billiard shape
parameter $\lambda$ for Robnik billiard with fixed density of boundary
discretization $b=12$. The numerical points are denoted by the diamonds, which
are joined by straight lines just to guide the eye.
\bigskip
\bigskip

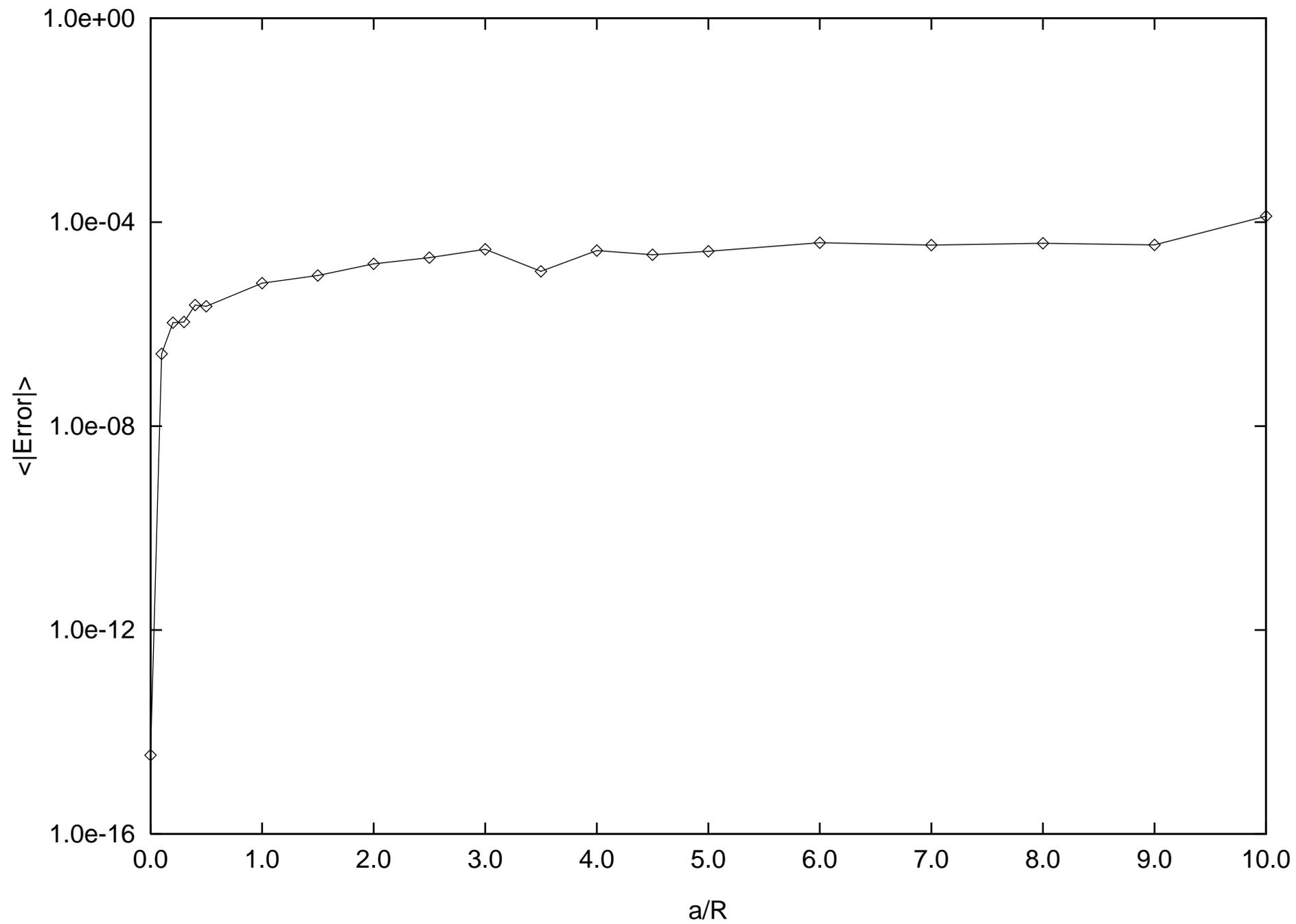
\noindent {\bf Figure 3:}
The ensemble averaged (over 100 lowest odd-odd eigenstates) absolute
error (measured in units of mean level spacing) versus the billiard shape
parameter $a/R$ for Bunimovich stadium with fixed density of boundary
discretization $b=12$. The numerical points are denoted by the diamonds, which
are joined by straight lines just to guide the eye.
\bigskip
\bigskip

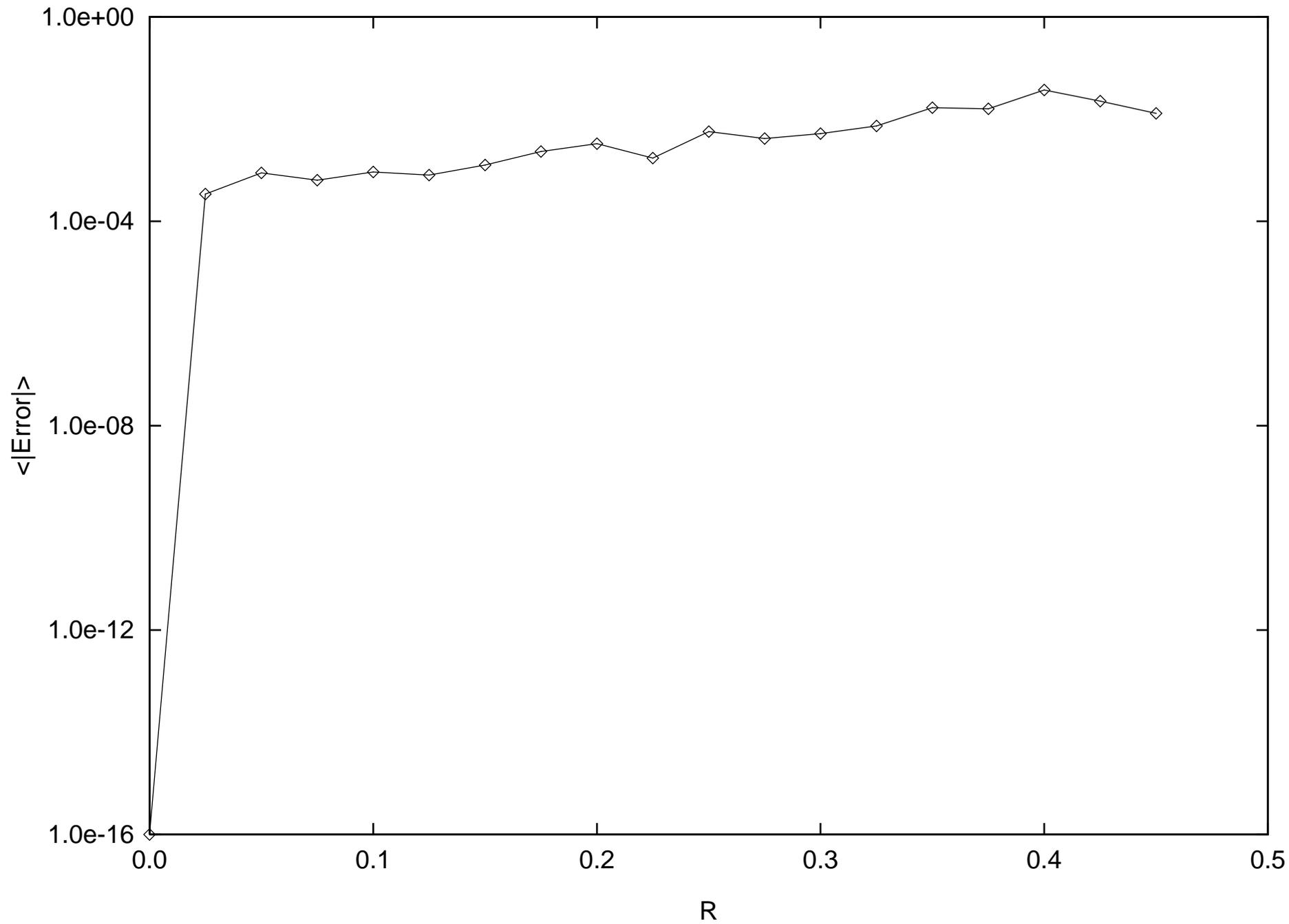
\noindent {\bf Figure 4:}
The ensemble averaged (over 100 lowest eigenstates)
absolute error (measured in units of mean level spacing)
versus the billiard parameter  $R$ (the radius of inner circle) for
the desymmetrized Sinai billiard with fixed density of boundary
discretization $b=12$. The numerical points are denoted by the diamonds, which
are joined by straight lines just to guide the eye.
\bigskip
\bigskip

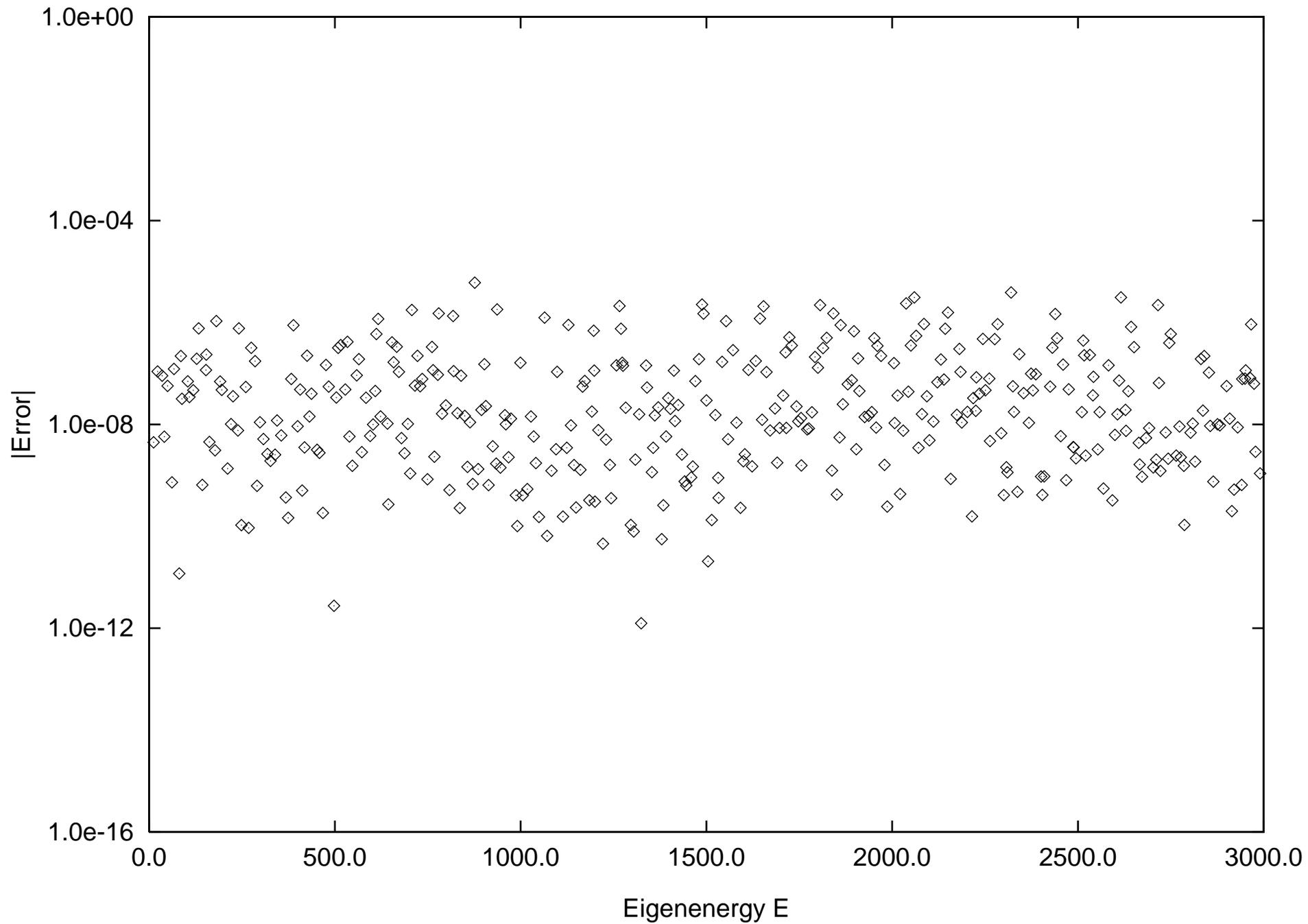
\noindent {\bf Figure 5:}
The absolute error of eigenstates versus eigenenergy for
the lowest 400 odd eigenstates of Robnik billiard at $\lambda=0.27$.
The averages over consecutive stretches (of 100 states each) are given in
table 2, demonstrating that \dE is quite stationary.

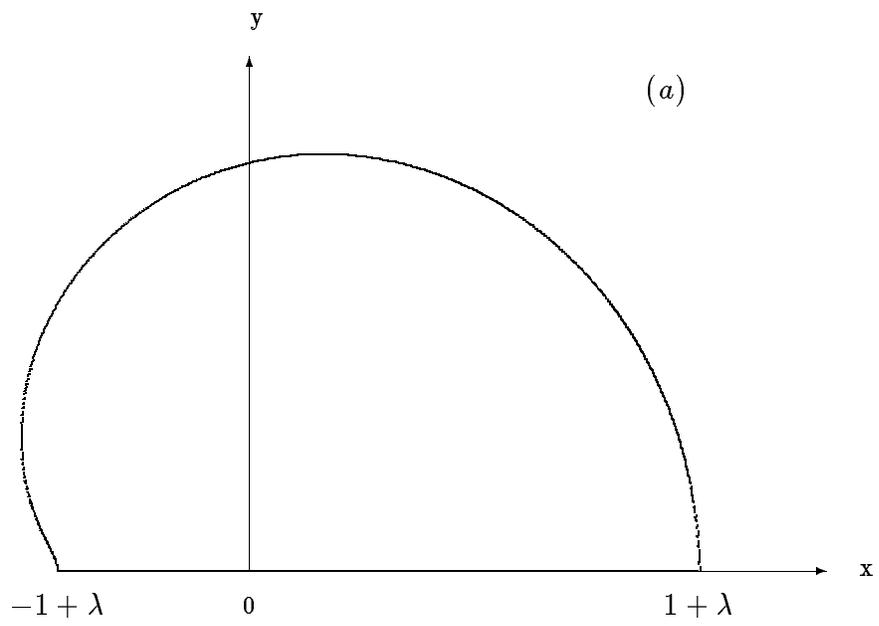
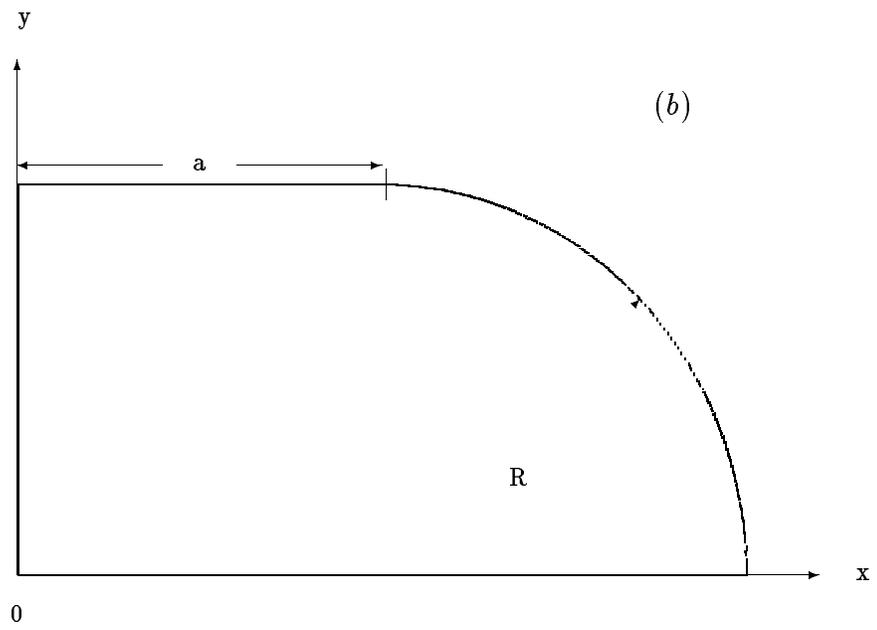
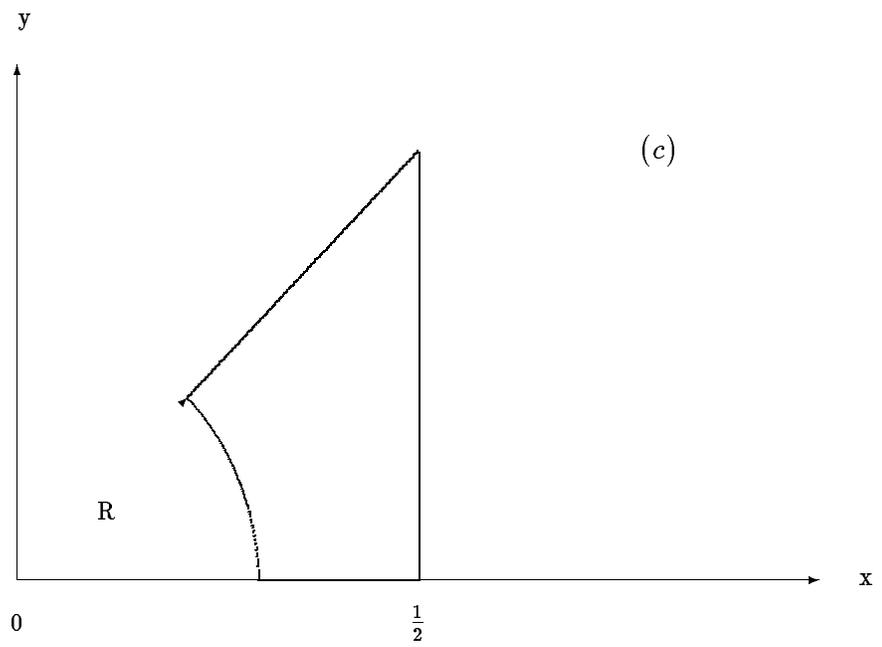
\end{document}